\documentclass[pre,reprint, superscriptaddress, nofootinbib, amsmath,amssymb, aps,longbibliography]{revtex4-1}

\usepackage{graphicx}
\usepackage{dcolumn}

\usepackage{bm}
\usepackage{caption}
\usepackage{float}
\usepackage{tikz-feynman}
\usepackage{lipsum}
\usepackage{amsmath}
\usepackage{mathtools}
\usepackage{amssymb}
\usepackage{dsfont}
\usepackage{color}
\usepackage{tikz}
\usetikzlibrary{automata, positioning, arrows}
\usepackage{nccmath}
\usepackage{caption}
\usepackage{subcaption}
\usepackage[utf8]{inputenc}
\usepackage{babel}
\usepackage{amsmath,amsthm,amssymb}
\usepackage{graphicx}
\usepackage{tikz}
\usepackage{tikz-feynman}
\tikzfeynmanset{compat=1.1.0}

\usepackage{physics}

\usepackage{listings}
\usepackage{enumitem}
\usepackage{microtype}
\usepackage[raiselinks=true, linktoc=all,hidelinks]{hyperref}
\usepackage[capitalise]{cleveref}

\newcolumntype{L}{>{$}l<{$}}

\allowdisplaybreaks[1]

\begin{document}
\title{A Unified Approach to Strong Local Correlations and Collective Fluctuations:\\ Eliminating Divergence in the Spin Channel} 

\begin{abstract}
Dynamical mean-field theory (DMFT) provides an optimal local approximation for correlated lattice systems by mapping the lattice onto a self-consistent effective impurity model. To account for the missing long-range correlations, we propose a novel extended approach, which we term fluctuating dynamical mean-field theory (fDMFT). It incorporates collective fluctuations of auxiliary impurity models across different sites via functional integration. Technically, this method involves obtaining a family of DMFT solutions on a grid for a self-consistent auxiliary classical field applied to the lattice. While the result can, in principle, be improved diagrammatically, we find that the minimal version of the theory already yields accurate results, with lowest-order diagrammatic corrections offering only minor improvements. This consistent framework, based on our fluctuating local field concept, demonstrates superior performance for the nearly half-filled Hubbard model compared to other known diagrammatic extensions of DMFT.

\end{abstract}

\author{S.D. Semenov}\email{roporoz@gmail.com}
\affiliation{Russian Quantum Center, Moscow 121205, Russia}
\affiliation{Moscow Center for Advanced Studies, 123592 Moscow, Russia}
\author{A.I. Lichtenstein}
\affiliation{Institut f{\"u}r Theoretische Physik, Universit{\"a}t Hamburg, Notkestra{\ss}e 9, 22607 Hamburg, Germany}
\author{A.N. Rubtsov}
\affiliation{Russian Quantum Center, Moscow 121205, Russia}
\affiliation{M.V. Lomonosov Moscow State University, Moscow 119991, Russia}
\date{\today}
 
\maketitle

\section{Introduction}
Mott–Hubbard insulators are perhaps the earliest and most prominent example demonstrating the importance of electron correlations in solids. Although band theory would predict these materials to be metallic, they instead become dielectric because of the spatial localization of electrons due to their Coulomb repulsion. Since the early days \cite{Mott_1968}, scientists have viewed local electron correlations at lattice sites as the primary mechanism leading to the formation of the Mott state. Later, the celebrated Dynamical Mean-Field Theory (DMFT) \cite{DMFT} was introduced to treat local correlations exactly. This approach indeed enables realistic modeling of bulk three-dimensional Mott insulators, such as transition‑metal oxides \cite{Kotliar_RMP_2006}. 

However, quasi‑two‑dimensional correlated systems described by the nearly half‑filled Mott–Hubbard model on a square lattice remain challenging. One of the main difficulties is the strong antiferromagnetic (AF) instability of the Fermi–Hubbard model near half‑filling \cite{DF_Rev}. Although the Mermin–Wagner theorem forbids true AF order in planar systems, this instability leads to strong collective spin fluctuations that cannot be handled within the local DMFT approach.

Currently, progress in the field of correlated Fermi–Hubbard systems is driven mainly by advances in “brute‑force” numerical methods, such as the tangent‑space tensor renormalization group (tanTRG) variant of the DMRG method \cite{Li2023} and the diagrammatic Monte Carlo approach using the so‑called connected determinant (CDet) scheme \cite{Rossi2017}. The latter can also be viewed as a “connected‑diagram” version of the standard continuous‑time QMC solver for a general impurity problem \cite{CTQMC}. 
The tanTRG method has enabled stable solutions for a $10\times10$
Hubbard cluster with open‑boundary conditions \cite{Li2023}, while the CDet approach has resolved the long‑standing problem of the pseudogap phase diagram of the Fermi–Hubbard model \cite{Georges2024}. One can also note experimental investigations of such systems using Fermi–Hubbard quantum simulators, such as the quantum gas microscope \cite{Bloch2026}.

Nonetheless, there has been almost no progress toward a simple scheme for describing strong nonlocal fluctuation phenomena based on the DMFT approximation \cite{DMFT}, one that would allow for an accurate and physically meaningful understanding of strong‑coupling nonlocal magnetic fluctuations.

Certain progress can be made using nonlocal DMFT extensions such as Dual Fermions (DF) \cite{Rubtsov_DF, Rubtsov_DB}, the Dynamical Vertex Approximation (D$\Gamma$A)\cite{DGA1, DGA2}, and the (dual) TRiply Irreducible Local EXpansion (TRILEX) \cite{D-TRILEX1, D-TRILEX2}, which incorporate nonlocal physics beyond purely local DMFT. These approaches often perform well in the weak‑to‑intermediate coupling regime and can capture short‑range antiferromagnetic correlations. However, in strongly correlated systems at low temperatures—where long‑range order and collective magnetic fluctuations dominate—these methods often become insufficient.

An alternative route is provided by parquet-type diagrammatic expansions beyond DMFT \cite{Astretsov2020, Kugler2025}. However, such schemes tend to become numerically unstable for strongly correlated systems at low temperatures, and their generalization to multi‑orbital systems is also quite complicated. We also refer to the extensive comparison~\cite{PRX_2021} of different methods applied to the half‑filled, moderately correlated Hubbard model with $U=2t$. 
In short, even for this relatively simple case, none of the existing approximations works reliably in the regime dominated by collective spin fluctuations—namely, well below the DMFT N{\'e}el transition point.

A reasonable pathway to describe magnetic fluctuations is to employ a scheme in which spin degrees of freedom are explicitly coupled to auxiliary fluctuating fields. However, attempts to introduce a path‑integral representation with fully quantum, noncollinear, rotationally invariant fluctuations \cite{Schulz1990, Lee1991, Siggia1988} have not yet produced numerical results due to the severe sign problem in lattice QMC calculations.

In this situation, the so‑called Fluctuating Local Field (FLF) scheme \cite{FLF_original_2016, Rubtsov_FLF2023, Yana_FLF2022, FLF_serirs, Savva_FLF2024, Silakov_FLF2026} was proposed as an optimal compromise between these approaches and simple mean‑field schemes. In the FLF ansatz, fluctuating fields are introduced in one or a few channels associated with the relevant collective fluctuations. Integration over the fluctuating field(s) is carried out explicitly, while the remaining degrees of freedom are treated approximately. Depending on the chosen approximation, different versions of the FLF scheme are obtained. In Ref. \cite{Rubtsov_FLF2023}, the moderately correlated Hubbard model was studied using FLF combined with a low‑order diagrammatic approximation.

In this paper, we introduce the FLF scheme based on the DMFT approximation and benchmark it using a strongly correlated system. Specifically, we present our approach for the single-band fermionic Hubbard model defined on an $L\times L$ periodic square lattice at inverse temperature $\beta$ in the presence of an external field ${\bf m}$.
We assume that ${\bf m}$ is conjugate to the leading mode associated with collective fluctuations.
This assumption places certain constraints on the system size $L$. The formation of a collective mode requires the system to contain a large number of particles, so $L$ must be sufficiently large. On the one hand, an assumption that fluctuations occur mostly in a single mode means that the system is not very large. On the other hand, assuming that fluctuations occur primarily in a single mode implies that the system should not be too large. In our practical calculations, we find that the theory performs well for $L=6$ to $10$. A more detailed discussion of the theory’s domain of applicability is provided after the presentation of these numerical results.

\section{Definitions}
\label{Definitions}

We use the Lagrangian formalism and tensorial notation, so the model of correlated lattice fermions with the bare energy spectrum ($\varepsilon$) in external fields (${\cal M}$, determined by the vector ${\bf m}$) is described by the imaginary-time action 
\begin{equation}\label{SHubbard}
    S_{\bf m}[c^\dag, c]=S^{at}[c^\dag, c]+\left(c^\dag \varepsilon c\right)-\left(c^\dag {\cal M}_{\bf m} c\right),
\end{equation}
where $c^\dag, c$ are Grassmann variables corresponding to the creation-annihilation operators.

The action of the atomic system reads:
\begin{equation}\label{Sat}
    S^{at}=-\left(c^\dag\left[-\frac{\partial}{\partial \tau} +\frac{U}{2}+\mu\right]c\right)+ \sum_j \int_0^\beta d\tau U n_{\tau,j, \uparrow} n_{\tau,j, \downarrow}. 
\end{equation}
It describes a lattice made of isolated atoms with the on-site repulsion $U$; the index $j$ numbers lattice sites, $\tau$ is the imaginary time and $n_{\tau,j, \sigma}=c^\dag_{\tau,j, \sigma} c_{\tau,j, \sigma}$.

The scalar product $\left(c^\dag \varepsilon c\right)$ can be explicitly written as 
\begin{equation}\label{S0explicit}
\left(c^\dag \epsilon c\right)=\sum_{\omega, {\bf k}, \sigma} \varepsilon_{\bf k} c^\dag_{\omega, {\bf k}, \sigma}  c_{\omega, {\bf k}, \sigma}    
\end{equation}
The indices run over the fermionic Matsubara frequencies $\omega=\pm \frac{\pi}{\beta}, \pm \frac{3 \pi}{\beta}, ...$, sites of the reciprocal lattice $k_{x, y}=-\pi + \frac{2 \pi }{L}, ..., \pi-\frac{2 \pi }{L}, \pi$, and spin projections $\sigma=\uparrow, \downarrow$, respectively.

The dispersion relation $\varepsilon_{\bf k}$ will be treated in its general form when presenting the formalism. To give numerical examples, we use   
\begin{equation}
\varepsilon_{\bf k}=-2 t (\cos k_x+\cos k_y),    
\end{equation} 
which describes nearest-neighbor coupling with the hopping amplitude $t$.  

The system will be considered at and near the half-filling. In this regime, fluctuations belong to the antiferromagnetic (AF) channel. Therefore it is relevant to consider the AF-ordering external field. The scalar product $\left(c^\dag {\cal M}_{\bf m} c\right)$ reads: 
\begin{equation}\label{AFfield}
    \left(c^\dag {\cal M}_{\bf m} c\right)=\beta N ({\bf m}{\bf s} ),
\end{equation}
where $\bf s$ is the AF polarization:
\begin{equation}\label{AFpolarization}
    {\bf s}=\frac{1}{\beta N}\int_0^\beta  d\tau \sum_{j\sigma \sigma '} e^{i \left({\bf Q} {\bf r}_j\right)}  c^\dag_{j \tau \sigma} {\bf \sigma}_{\sigma \sigma '}  c_{j \tau \sigma'}   \, ,
\end{equation}
where ${\bf \sigma}=(\sigma^x, \sigma^y, \sigma^z)$ is the Pauli vector, ${\bf r}_j$ are positions of the sites, and the AF-vector ${\bf Q}=(\pi,\pi)$.

The key concept of the Dynamical Mean-Field Theory (DMFT)~\cite{DMFT} is an approximation related to a lattice of so-called Anderson impurity models. Later we use the short notation $\cal D$ and ${\cal I}$ for DMFT and Impurity, respectively.  This is essentially a set of atoms, each of which has been placed in an effective bath $\Delta_j$. 
The corresponding action is 
\begin{equation}\label{Simpurity}
    S^{\cal I}_{\bf m}[c^\dag, c]=S^{at}[c^\dag, c]+\left(c^\dag \Delta c\right)-\beta N ({\bf m} {\bf s}),
\end{equation}
where $\left(c^\dag \Delta c\right) = \sum_{\omega, j, \sigma, \sigma'} c^\dag_{\omega, j, \sigma} \Delta_{\omega, j, \sigma, \sigma'} c_{\omega, j, \sigma'}$. Note that 
this action describes a set of sites which do not talk to each other. 
Each of them is immersed in its own bath $\Delta_j$ and therefore can be treated separately from others. This makes the impurity system (\ref{Simpurity})  much simpler than the original lattice model (\ref{SHubbard}).

In the DMFT, the hybridization is $\omega$-dependent, hence explain a term ``dynamical'' in the name of the method. This does not allow for an analytical treatment of the impurity system.
Nevertheless, there are effective numerical solvers for the impurity problem~\cite{DMFT,CTQMC}.  Thus, given the hybridization function, one can obtain the single-particle Green's function of the impurity system $G^{\cal I}_{12}=-\langle c_{1} c^\dag_{2}\rangle$. This quantity is diagonal in $j$-indices, as well as the self-energy defined by the Dyson equation $\Sigma^{\cal I}=i\omega + {\cal M}_{\bf m} -\Delta  -G_{\cal I}^{-1}$. 

In a nutshell, the DMFT assumes that the lattice self-energy function can be approximated with $\Sigma^{imp}$, so that 
the approximate lattice Green's function equals
\begin{equation}
\label{G_DMFT}
    G^{\cal D}_{\bf m}=\frac{1}{(G^{\cal I}_{\bf m})^{-1}+\Delta_{\bf m}-\varepsilon}.
\end{equation}

As a condition for the hybridization $\Delta$, one requires that the local part $G^{loc}$ of the lattice Green's function $G^{\cal D}$ coincides the impurity one,
\begin{equation}\label{DMFT_SC}
    G^{\cal I}_{\bf m}=G^{loc}_{\bf m};
\end{equation}
the equality holds for all values of the frequency and spin indices. A practical solution of this equation requires performing the famous DMFT iterative loop~\cite{DMFT}.

\begin{figure}[!h]
  \centering
  \includegraphics[width=\columnwidth]{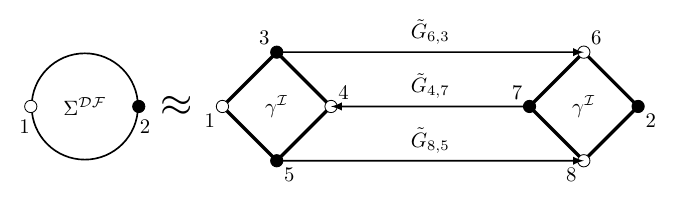}
  \caption{A non-local Dual Fermion (DF) leading self-energy diagram $\Sigma_{12}^{\cal DF} (i \omega)$ built from two local impurity irreducible vertices $\gamma^{\cal I}$ connected by three dual propagators $\tilde G$. The dual propagator is defined as $\tilde{G}_{k, l}=G^{\cal D}_{k, l}-G^{\cal I}_{k, l}$, where $k$ and $l$ are generalized indices collecting the node label (or momentum), Matsubara frequency (or imaginary time), and spin, e.g. $k\equiv(j,i\omega_n,\sigma)$. Repeated internal indices are summed over, while the external indices are fixed.}
  \label{fig:DF_diag}
\end{figure}

To finalize this section, let us also outline formulas for the Dual Fermion (DF) extension of the DMFT scheme, later labeled ${\cal DF}$.  The DF result for Green's function can be written in the form
\begin{equation}\label{G_DF}
    G^{\cal DF}_{\bf m}=\frac{1}{(G^{\cal I}_{\bf m} + G^{\cal I}_{\bf m} \Sigma^{\cal DF}_{\bf m} G^{\cal I}_{\bf m})^{-1}+\Delta_{\bf m}-\varepsilon},
\end{equation}
where $\Sigma^{\cal DF}$ is so-called dual fermion self-energy. This quantity can be expanded in diagram series. 
The leading term in this series is the second-order diagram involving a pair of four-point vertices and three dual Green’s functions, as shown in Fig. \ref{fig:DF_diag}.
The four-point vertex $\gamma^{\cal I}$ 
corresponds to the effective two‑particle interaction of the impurity model, while the bare lines represent the nonlocal part of the DMFT Green’s function or a DF-propagator, 
${\tilde {G}}=G^{\cal D}-G^{\cal I}$.  
As a result, the leading contribution to $\Sigma^{\cal DF}$ is fully nonlocal in space.

More sophisticated consideration can include higher-order and bold-line (skeleton) diagrams for $\Sigma_{\cal DF}$, as well as possible corrections to the hybridization function $\Delta$. 
We refer to the papers \cite{Rubtsov_DF, Rubtsov_DB}, for a more detailed description of the Dual Fermion (DF) technique, as well as for the explicit algebraic expressions corresponding to the diagrams.

\section{Derivation of \lowercase{f}DMFT and \lowercase{f}DF}

Let us start with the partition function 
\begin{equation}\label{Partition}
Z_{\bf m}=\int D[c c^\dag] \, e^{-S_{\bf m}[c^\dag, c]}  \, ,     
\end{equation}
and use a following representation of the unitary operator
\begin{equation}
    1=\left(\frac{\lambda}{2 \pi \beta N}\right)^{3/2} \int e^{-\frac{\beta N}{2\lambda}({\bf h}-{\bf m}-\lambda {\bf s})^2} d^3 h,
\end{equation}
where $\lambda$ is a positive parameter, and the vector $\bf h$ will be referred to as fluctuating field. 

We insert this unitary into the integrand of the partition function expression (\ref{Partition})and rewrite the resulting equation as follows:
\begin{equation}\label{FLFensemble}
    Z_{\bf m}=\left(\frac{\lambda}{2 \pi \beta N}\right)^{3/2} \int \tilde Z_{\bf h} e^{-\frac{\beta N}{2\lambda} ({\bf h}-{\bf m})^2}  d^3 h, 
\end{equation}
where  
\begin{equation}\label{Partition.nu}
    \tilde Z_{\bf h}=\int D[c c^\dag] \, e^{-\tilde S_{\bf h}[c^\dag, c]}  
\end{equation}
with 
\begin{equation}\label{Action.nu}
    \tilde S_{\bf h}[c^\dag, c]=S^{at}[c^\dag, c]+\left(c^\dag \varepsilon c\right) - \beta N ({\bf h} {\bf s}) +  \frac{\lambda}{2} \beta N s^2.
\end{equation}
Later, we will also use the specific Fluctuating Local Field (FLF) free energy $\tilde F_{\bf h}$ defined by the formula
\begin{equation}
    \beta N \tilde F_{\bf h} = -\ln\tilde Z_{\bf h}.
\end{equation}
The quantity $\beta N \tilde F_{\bf h}$ acts as an effective potential for the field $\bf h$.

This way, we presented the original system (\ref{Partition}) as an ensemble of lattices (\ref{FLFensemble}), each subjected to some fluctuating field ${\bf h}$. Compared to the original action $S$, the modified one $\tilde S$ includes an additional counter-AF interaction that is fully nonlocal in space and time. It means that for each of the systems (\ref{Partition.nu}) fluctuations are suppressed, compared to the original one (\ref{Partition}). Since the transformations are exact so far, these fluctuations are not missed: they are to be restored after the integration over the ensemble (\ref{FLFensemble}).

Later, we will be interested in the properties of the system in the limit $m\to 0$ limit.
The Green's function can be found by the log-variation of (\ref{FLFensemble}):
\begin{equation}\label{Gav}
    G=\overline {\tilde G_{\bf h}},
\end{equation}
where  $\overline {A_{\bf h}}= Z^{-1}\int  A_{\bf h} \tilde Z_{\bf h} e^{-\frac{\beta N}{2\lambda} h^2}  d^3 h$, and $\tilde G_{\bf h}$ is the Green's function corresponding to the action $\tilde S_{\bf h}$.

Importantly, although each $\tilde G_{\bf h}$ remains local in the DMFT sense, the functional integration over the field ${\bf h}$ effectively reintroduces nonlocal physics: it captures fluctuations of the leading nonlocal mode and feeds their impact back into observables through the ensemble average~\eqref{Gav}. This observation also explains our terminology fluctuating Dynamical Mean Field Theory (fDMFT): instead of a single self-consistent DMFT solution, the theory generates a whole ensemble of DMFT impurity problems parametrized by ${\bf h}$, whose solutions fluctuate with the field and are then averaged with the FLF weight.

Given the Green's function, we can obtain the self-energy from its definition: $\Sigma=i\omega -\epsilon- G^{-1}$. At this point, we emphasize that while the Green's function is a linear average of $\tilde G_{\bf h}$, the self-energy   is averaged in an essentially nonlinear way. In particular, we will see that this nonlinearity leads to a $\bf k$-dependent $\Sigma$ even if we assume that each $\tilde S_{\bf h}$ obeys a fully local self-energy.

One can also find the Curie constant $C$. Using the formula $C=\left.\frac{1}{3 \beta^2 N} \Delta_{m} \ln Z\right|_{m=0}$, one obtains
\begin{equation}\label{Cav}
C=\frac{N \overline {h^2}}{3 \lambda^2} -\frac{1}{\beta \lambda}.
\end{equation}

The idea behind the FLF ansatz is  to consider each of the systems (\ref{Partition.nu}) approximately, and then perform a numerical integration over ${\bf h}$. It is expected that approximate methods of the mean-field family work better for  (\ref{Partition.nu}) than for (\ref{Partition}), because the former exhibits smaller fluctuations due to the ``extra'' counter-AF term. In our previous work we have developed the FLF scheme based on the diagrammatic expansion for (\ref{Partition.nu}). Such an approach works well for small and moderate correlations. Here we present a similar approach, but use a DMFT based series, thus aiming for the description of strongly correlated systems.

Using a mean-field decoupling of the long-range term in $\tilde S_{\bf h}$, we reduce it to the Hubbard lattice subjected to a self-consistent field. 
For this purpose, let us rewrite (\ref{Action.nu}) with the  average spin polarization $\tilde {\bf s}_\nu$ as follows:
\begin{equation}\label{tildeSimp.2}
     \tilde S_{\bf h}[c^\dag, c]=S_{{\bf h}-\lambda \tilde {\bf s}}[c^\dag, c]- \frac{\lambda}{2} \beta N  \left(\tilde {\bf s}_{\bf h}^2- ({\bf s}-\tilde {\bf s}_{\bf h})^2 \right).
\end{equation}
In this equation, the term $({\bf s}-\bar {\bf s}_{\bf h})^2$ describes collective spin fluctuations. These fluctuations are suppressed by the counter-AF interaction.
Additionally we note that fluctuations are inverse proportional to the number of sites. 
 
This reasoning justifies the mean-field approximation in which the term $({\bf s}-\bar {\bf s}_{\bf h})^2$ is neglected.
After this simplification, $\tilde S_\nu$ describes the Hubbard lattice subjected to the following self‑consistent field: 
\begin{equation}\label{effective.field}
   \tilde {\bf m}={\bf h}-\lambda \tilde {\bf s}_{\bf h}.
\end{equation}
Therefore, the Green's function of the lattice (\ref{Action.nu}) can be approximated with 
the DMFT expression (\ref{G_DMFT}) at this field,
\begin{equation}\label{G.via.GDMFT}
    \tilde G_{\bf h}=G^{\cal D}_{\tilde {\bf m}}.
\end{equation}
and, consequently,
\begin{equation}\label{s2s}
    \tilde s_{\bf h}= \bar s^{\cal D}_{\tilde {\bf m}}.
\end{equation}
The equations (\ref{effective.field}) and (\ref{s2s}) allow a self-consistent calculation of $\tilde s_{\bf h}$.

The Fluctuating Local Field estimation for specific free energy $\tilde F_{\bf h}$ and, therefore, the partition function $\tilde Z_{\bf h}$ can be found by integrating of the differential identity 
\begin{equation}
   \beta N d \tilde{F}_{\bf h}=(\bar s_{\bf h} d{\bf h})-\frac{d h}{\lambda}
\end{equation}

This equation can be integrated numerically, using the DMFT  value for $\bar s_{\bf h}$.  
The Green's function and Curie constant are given by the equations (\ref{Gav}), (\ref{Cav}), and (\ref{G.via.GDMFT}). These expressions constitute our fDMFT scheme.

\begin{figure}[t]
  \centering
  \includegraphics[width=\columnwidth]{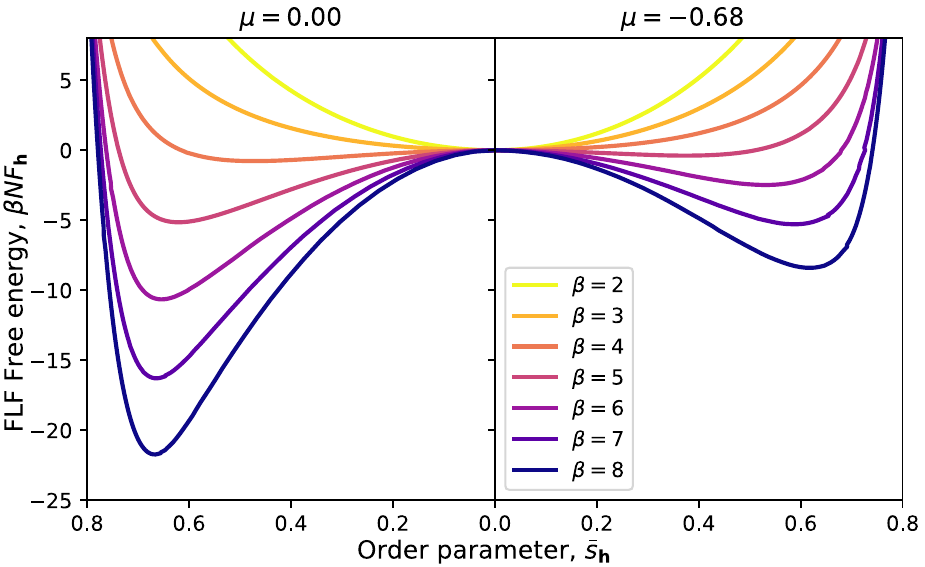}
  \caption{The Fluctuating Local Field (FLF) free energy $\tilde{F}_{\bf{h}}$ as a function of the order parameter $\bar{s}_{\bf{h}}$ for the square cluster with size $L = 8$ calculated via fDF approach.
  Left panel: $\mu=0.00$; right panel: $\mu=-0.68$.
  Colors indicate different inverse temperatures $\beta$.}
  \label{fig:free_energy_s}
\end{figure}

As seen in Fig.~\ref{fig:free_energy_s}, the effective potential $\beta N \tilde F_{\bf h}$ acquires a characteristic ''Mexican-hat'' shape well known from Landau theory \cite{Landau_PT, Savva_FLF2024}. The emergence of a pronounced minimum at finite $\bar {\bf s}_{\bf h}$ indicates the formation of a collective antiferromagnetic mode, while the finite width of the minimum reflects the presence of fluctuations in the system. As expected, at elevated temperatures and upon doping the antiferromagnetic mode is progressively suppressed, which is manifested by the disappearance of the minimum and a crossover to a single-well potential centered at $\bar {\bf s}_{\bf h}=0$.

Possible further improvement of the fDMFT relies on corrections to 
the expression (\ref{G.via.GDMFT}) because all other transformations are exact.
The expression (\ref{G.via.GDMFT}) is based on two approximations: mean-field decoupling of the effective long-range interaction and subsequent use of the DMFT scheme.
As we discussed, the first one is valid because collective spin fluctuations in $\tilde S$ are small. The DMFT approximation can be improved by replacing $G^{\cal D}$ with $G^{\cal DF}$ in (\ref{G.via.GDMFT}). This provides us with the fluctuation Dual Fermion (fDF) method. Later we present the fDF results obtained  with a single DF diagram depicted in Fig. \ref{fig:DF_diag}.

Value of parameter $\lambda$ remains undefined so far. In our previous papers, we have found that $\lambda=\frac{U}{2}$ works well for weakly and medium-correlated systems.
However it is not a priori clear if this value remains a good choice for a strongly correlated ensemble. Here we consider how to choose $\lambda$ in a generic case.

As we pointed out, $\lambda$ should be adjusted to suppress collective fluctuations in the AF mode. Formally this can be interpreted as a requirement that the AF susceptibility of the lattice model (\ref{Action.nu}) is the same as that of the uncorrelated system with the same Green's function, $\tilde \chi= \chi_{0}$, where $\chi_{0}= -\frac{1}{\beta N} \sum_{k, \omega} G_{\omega, {\bf k}} G_{\omega, {\bf k}+{\bf Q}}$. The long-range counter-AF interaction adds an additional contribution to the polarization operator, compared to the DMFT susceptibility: $\tilde \chi^{-1}=\chi_{\cal D}^{-1}+\lambda$. These two formulas provide the desired criterion for the optimal value of $\lambda$ which we mark with an asterisk:
\begin{equation}\label{lambda}
    \lambda^{*}=\chi_{0}^{-1}-\chi_{\cal D}^{-1}. 
\end{equation}
In our practical calculations, we obtain $\lambda$ by estimating the r.h.s. of this equation at $m=0$. 

The susceptibility is calculated numerically as the following derivative
\begin{equation}
\chi_{\cal D} = \frac{d\bar s^{\cal D}}{dm}.
\end{equation}
 We note that the susceptibility $\chi_{\cal D}$ diverges at the DMFT N\'eel point and is negative at lower temperatures, whereas the {\it exact} spin susceptibility of a finite system is always finite-positive because fluctuations suppress the phase transition to the ordered phase. Our FLF approach will reproduce this behavior correctly.

We found that the value of $\lambda$ obtained with our procedure goes to $\frac U2$ at small $U$. Moreover, for the system with $U=5 t$ we typically obtain $\lambda^*\approx 0.4 U$, and we have checked numerically that quantities (\ref{Gav}) and (\ref{Cav}) in fact depend on the value of $\lambda$ quite weakly. Therefore, $\lambda^*=\frac{U}{2}$ seems to be a reasonable option for strongly correlated systems as well as for moderate-correlated ones. Nevertheless, we consider (\ref{lambda}) to be the best justified choice in a general case.

Having established the fDMFT formalism in the previous sections, we now turn to its numerical implementation and the resulting physical observables. The purpose of this section is to present the numerical results and provide their physical interpretation.

\section{Numerical results}
Throughout this section we consider the Hubbard model with the Hamiltonian introduced above. Unless stated otherwise, we fix a two-dimensional periodic square lattice of size $L = 8 $ and study temperatures in the inverse-temperature range $\beta\in[0.5,10]$. One of the key advantages of the fDMFT (fDF) approach is that it can be applied straightforwardly away from half-filling; to illustrate this capability, we present results for $\mu=0$ and $-0.68$, covering the half-filled case and one example doped regime.

As a reference (benchmark) method we employ the determinant Quantum Monte Carlo (DQMC) results for assessing the accuracy of the fDMFT results and for identifying systematic trends across temperature and doping.

\subsection{Calculation procedure}

\begin{figure}
    \includegraphics[width=.97\columnwidth]{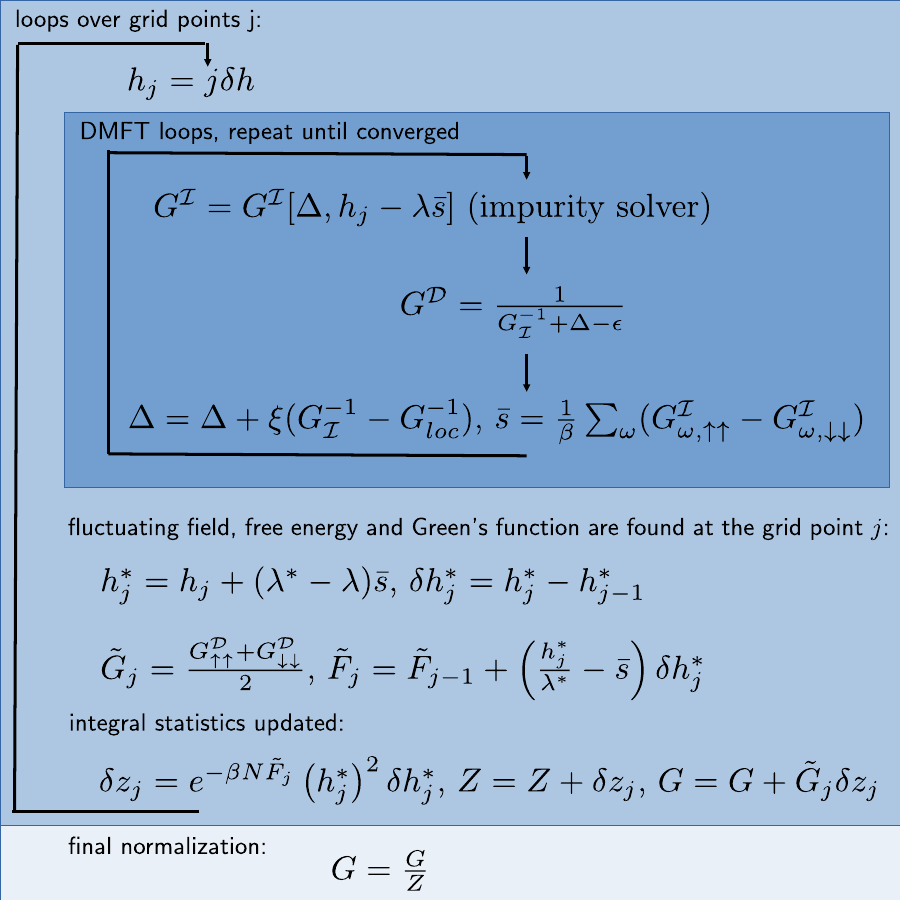} 
    \caption{Schematic DMFT+FLF calculation procedure. For each value of the fluctuating field $h_j=j\,\delta h$ we run an inner DMFT self-consistency loop until convergence. At each grid point we then compute the effective field $h_j^*$, the free-energy increment $\tilde F_j$, and the averaged Green's function $\tilde G_j$; the statistical weights $\delta z_j$ are accumulated to obtain $Z$ and $G$, followed by the final normalization $G=G/Z$.}\label{calculation}
\end{figure}

Let us explain  the implementation of the method in more practical details. For clarity, we describe how the fDMFT Green's function is calculated. Calculation of e.g. Curie constant can be implemented in the same scheme (\ref{Cav}). An account of the fDF correction is also straightforward, although the calculation of the impurity vertices requires some extra computing time.  

Integration over the $3$-component fluctuation field is greatly simplified because 
we consider the rotation-invariant system. In this case, the integration over angles is trivial: the partition function $\tilde Z_{\bf h}$ does not depend on angular variables, 
and angular integration of the Green's function $\tilde G_{\bf h}$ just gives its singlet part $\frac12(G_{\uparrow\uparrow}+G_{\downarrow\downarrow})$. The numerical procedure is therefore reduced to grid integration over a single variable $h$.

Preliminary, the integration over $h$ can be organized as follows. Consider the grid $h_j=j \delta h$ with some small step $\delta h$. At each grid point, one can perform DMFT loops with additional self-consistency included. The standard DMFT update for the hybridization function is 
\begin{equation}
    \Delta |_n = \Delta|_{n-1} +\xi (G_{\cal I}^{-1}-G_{loc}^{-1})|_{n-1},
\end{equation}
where $n$ numbers iterations and $0<\xi<1$ is a positive parameter adjusted to ensure the best convergence. In our case, one should vary also the acting AF field:
\begin{equation}
    \tilde m|_{n}=h_j-\lambda \bar s|_{n-1}.
\end{equation}
The fixed point of this iterative process is the DMFT solution for a proper self-consistent field (\ref{effective.field}).

The outlined method should be improved because of convergence issues. For the system studied, we observed that for $\lambda \approx 0.5 t$ the iterations show excellent converge within a broad range of parameters, including temperature well below the DMFT Néel point. On the other, the optimized value $\lambda^* \approx \frac{U}{2}$ leads to a divergency. Fortunately this issue can be easily cured. The key observation following from (\ref{effective.field}) is that, given the set of DMFT solutions for some $\lambda$ at the grid $h_j$, we are also provided with a solution for $\lambda^*$ at the grid 
\begin{equation}
h^*_j=h_j+(\lambda^*-\lambda) \bar s.    
\end{equation}

In our calculations, we perform iterations at grid points $h_j=j \delta h$ for $\lambda=0.5 t$ and then switch to a proper $\lambda^*$ obtained from (\ref{lambda}).
Our numerical scheme is illustrated by Fig.~\ref{calculation}.

\subsection{Self-energy $\Sigma_{\bf{k}} (iw)$}
To benchmark the ability of the our method to reproduce single-particle properties, we compare the self-energy $\Sigma_{\bf{k}}(i\omega_0)$ on the first Matsubara frequency $\omega_0=\pi/\beta$ caculated via fDF method (as more precise one), against the DQMC reference data. The lowest Matsubara frequency provides access to the essential low-energy physics: it is most sensitive to the renormalization and damping of quasiparticles near the Fermi surface, which ultimately control the development of ordering tendencies and the corresponding instabilities.

\begin{figure}[t]
  \centering
  \includegraphics[width=\columnwidth]{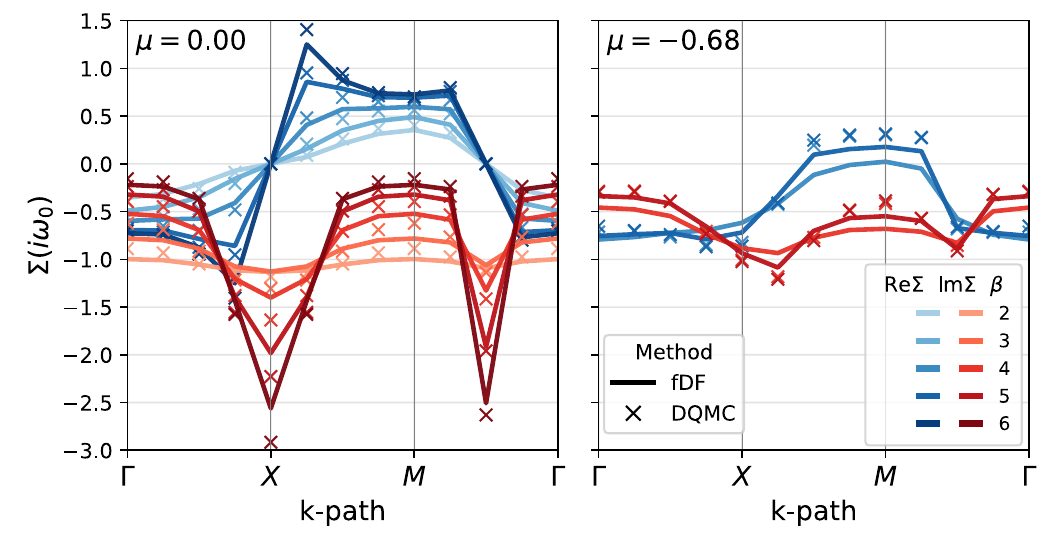}
  \caption{Self-energy at the first Matsubara frequency, $\Sigma(\mathbf{k},i\omega_0)$ with $\omega_0=\pi/\beta$,
  along the $\Gamma\!-\!X\!-\!M\!-\!\Gamma$ path.
  Lines: $fDF$, crosses: DQMC.
  Left: half-filling, $\mu=0.00$; Right:  doped case, $\mu=-0.68$.
  Data are shown for $\beta=2$--$6$ (as indicated in the legend).}
  \label{fig:sigma_w0}
\end{figure}

As shown in Fig.~\ref{fig:sigma_w0}, the fDF results for the self-energy closely reproduce the DQMC data throughout the temperature range $\beta=2$-$6$, both at half filling ($\mu=0.00$) and in the doped case ($\mu=-0.68$). Importantly, beyond the overall quantitative agreement, the fDF curves follow the dependence as DQMC along the $\Gamma\!-\!X\!-\!M\!-\!\Gamma$ path: the characteristic shape and its evolution with temperature are captured correctly. This demonstrates that the method reliably describes the essential single-particle physics in this parameter regime. 

One can also see that the numerical agreement gradually decreases as $\beta$ increases (lowering the temperature). This trend can be interpreted as a limitation of the present implementation, where we focused on only the main AF fluctuation mode. At lower temperatures, additional nonlocal fluctuations close to ${\bf Q}=(\pi,\pi)$ arise and are not included here.

\subsection{Antiferromagnetic susceptibility $\chi_{AF}$}
\label{sec:susceptibility}
To characterize many-body observables within the fDMFT (fDF) framework developed in this work, we consider the susceptibility in the AF spin channel. In the parameter regime close to half-filling ($n\approx 1$), antiferromagnetic correlations typically remain among the leading collective modes. At the same time, conventional single-site DMFT treatment may exhibit an artificial finite-temperature second-order transition to an AF-ordered state, manifested by an anomalously large (formally divergent) response to an external staggered field $\boldsymbol{m}$. Here we demonstrate that the susceptibility computed within the fDMFT (fDF) formalism provides a more realistic behavior: it captures the essential physics of AF fluctuations and is numerically close to the DQMC results, which we use as a reference.

\begin{figure}[h!]
  \centering
  \includegraphics[width=\columnwidth]{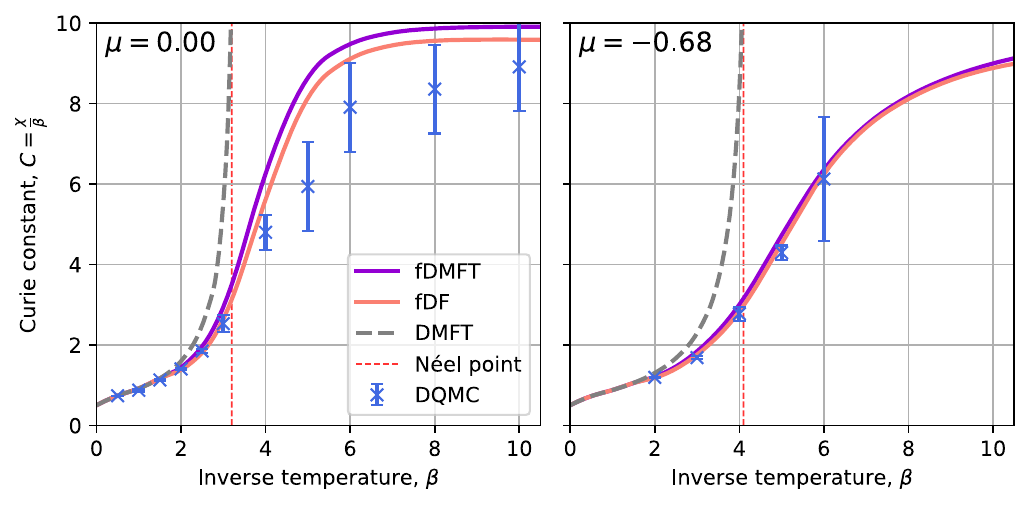}
  \caption{Curie constant $C=\chi/\beta$ in the antiferromagnetic channel as a function of inverse temperature $\beta$.
  Lines show DMFT, fDMFT, and fDF results; crosses indicate the DQMC reference data.
  Left panel: half filling, $\mu=0.00$; right panel: doped case, $\mu=-0.68$.
  The vertical dashed line marks the N\'eel point (see text).}
  \label{fig:chi}
\end{figure}

\begin{figure*}[!t]
  \centering
  \includegraphics[width=\textwidth]{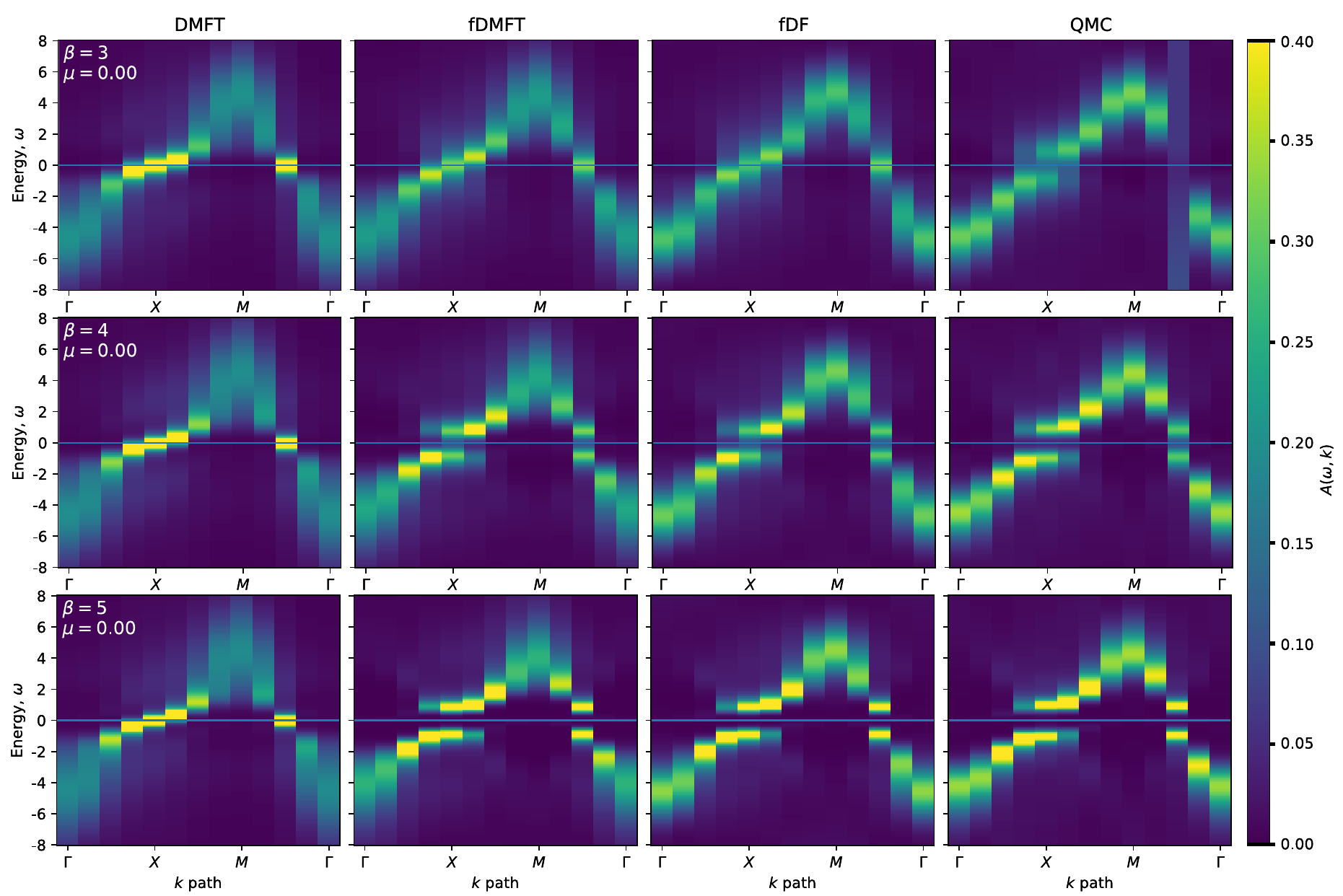}
  \caption{Momentum-resolved spectral function $A(\mathbf{k},\omega)$ on an $L =8$ square periodic cluster along the
  $\Gamma\!-\!X\!-\!M\!-\!\Gamma$ path. Columns correspond to the methods (DMFT, fDMFT, fDF, DQMC) and rows to
  inverse temperatures $\beta=3,4,5$ (as indicated on the panels). The spectral functions were obtained by
  analytic continuation of the Matsubara Green’s function~\cite{Anacont}.}
  \label{fig:Akw_matrix}
\end{figure*}
As can be seen in Fig.~\ref{fig:chi}, the susceptibility of AF obtained within the fDMFT (fDF) formalism is in excellent agreement with the QMC data over the entire temperature range shown. The remaining small deviations can be attributed to several systematic effects. 
First, in the present FLF implementation we retain only the leading AF collective mode, whereas including additional subleading modes would further improve the quantitative accuracy.

Second, the fDMFT (fDF) results inherit the finite numerical precision of the underlying DMFT impurity solver (e.g., discretization and statistical errors, as well as convergence tolerances of the impurity problem). 
Finally, the DQMC reference data are also subject to their own limitations, including finite-size effects and statistical uncertainties, which become more pronounced at low temperatures and especially strong-correlation regimes.

However, we highlight that fDMFT/fDF approach has several key practical advantages over DQMC. 
First, it remains reliable in the doped regime, where DQMC is severely limited by the fermionic sign problem, which leads to rapidly growing statistical noise and large error bars. 
Second, fDMFT/fDF can be applied to substantially larger inverse temperatures $\beta$, providing access to the low-temperature behavior of the Curie constant in a range that is typically out of reach for DQMC.

Importantly, the resulting $C(\beta)$ curves reveal a transparent physical trend. 
Upon cooling, AF correlations build up and the system enters a regime dominated by slowly fluctuating (quasi-ordered) staggered magnetization. 
Correspondingly, the Curie constant saturates and becomes nearly $\beta$-independent, signaling the emergence of a well-formed collective AF mode rather than a  high-temperature Curie growth.

\subsection{Spectral density function $A_{\bf{k}}(\omega)$}

\begin{figure*}[!t]
  \centering
  \includegraphics[width=\textwidth]{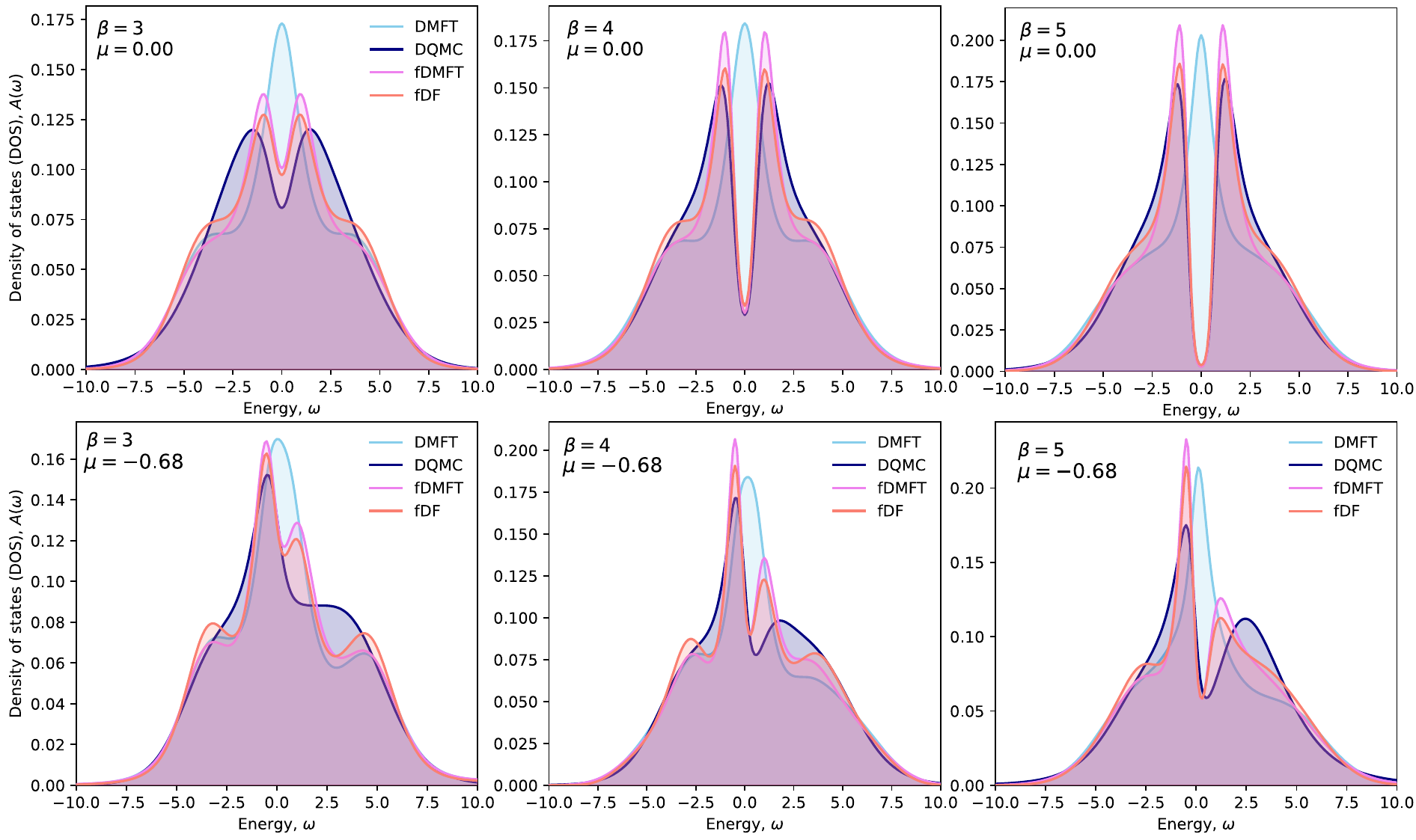}
  \caption{Density of states (DOS) $A(\omega)$ obtained by analytic continuation of the Matsubara Green’s
  function~\cite{Anacont}. Each panel shows all four methods (DMFT, fDMFT, fDF, and DQMC).
  Columns correspond to inverse temperatures $\beta=3,4,5$ (left to right), while rows correspond to chemical
  potentials (top: $\mu=0.00$; bottom: $\mu=-0.68$)}
  \label{fig:dos_wall}
\end{figure*}
The spectral function $A(\omega, \mathbf{k})$ for correlated lattice systems provides important information on the interplay between band-like dispersion and Mott--Slater physics. Our periodic $8\times 8$ system already has enough $\mathbf{k}$-resolution to see all these important features. We used the standard analytic continuation of the Green's function to the real frequency axis~\cite{Anacont}.

We first discuss a simple case of the half-filled Hubbard model for temperatures close to the Ne\'el point ($\beta \approx 3$), i.e., the undoped case.
Figure~\ref{fig:Akw_matrix} shows the results of the paramagnetic DMFT scheme as well as those of fDMFT and fDF, compared to the exact numerical sign-free DQMC scheme. One can see that the DMFT spectral function represents a metallic-like behavior and is very different from all other schemes. Moreover, the high-energy Hubbard bands are separated from the quasi-particle bands which cross the Fermi level. Already the fDMFT scheme drastically changes this situation and produces an insulator-like spectral function with a gap near the $X$-point in the Brillouin zone as well as at $\Gamma M/2$. This feature is related to an antiferromagnetic-like Slater splitting and coexists with non-dispersive high-energy Hubbard bands. A more elaborate fDF scheme, which improves the $\mathbf{k}$-dependence of correlation effects, shifts the spectral weight near the $\Gamma$ point towards lower-lying Hubbard states from the DMFT quasi-particle bands, which results in a reduction of the strong incoherent features in fDMFT.
This makes the results of fDF very close to the reference DQMC spectral function, with a small underestimation of the Slater splitting near the Fermi level.

\begin{figure*}[!t]
  \includegraphics[width=\textwidth]{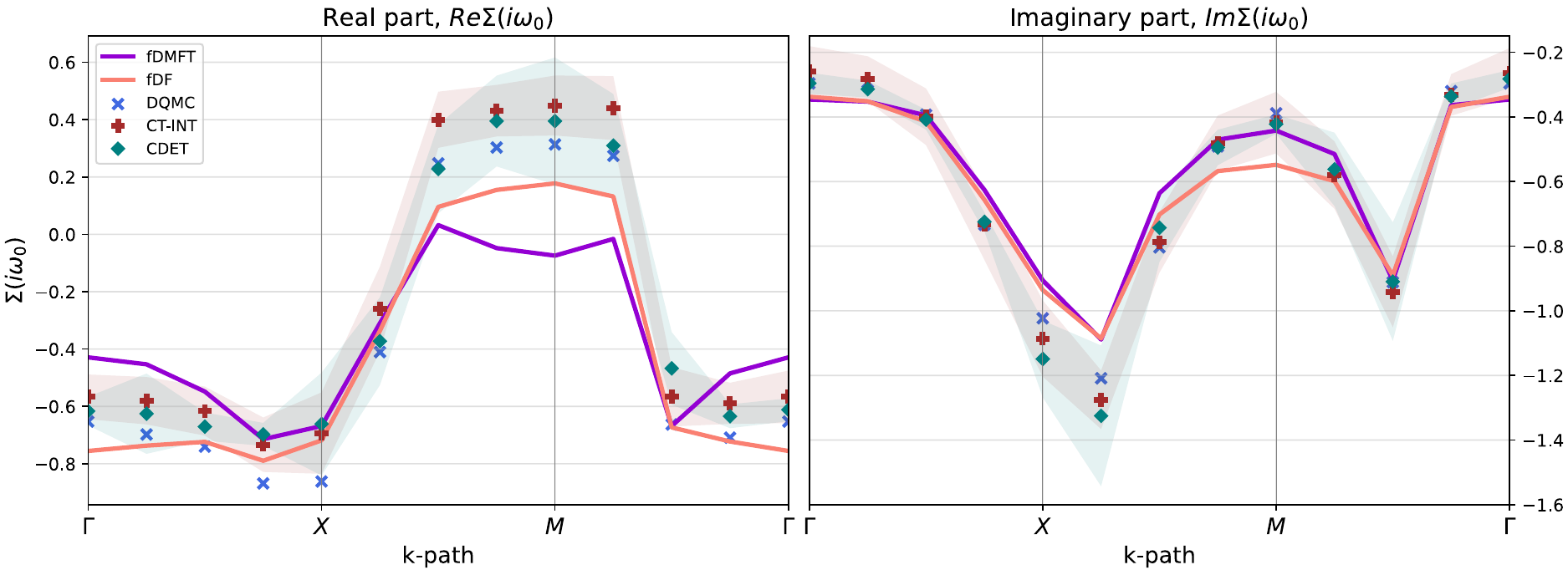}
  \centering
  \caption{Self-energy at the lowest Matsubara frequency, $\Sigma_{\mathbf{k}}(i\omega_0)$ with
  $\omega_0=\pi/\beta$, along the $\Gamma\!-\!X\!-\!M\!-\!\Gamma$ path for the $L = 8$ square periodic Hubbard cluster at
  $U=5$, $\beta=5$, and $\mu=-0.68$ ($\delta n=0.06$).
  Solid lines show the fDMFT and fDF results, while symbols denote the reference data from DQMC, CDet, and CT-INT \cite{Georges2024}. Shaded regions indicate the statistical uncertainty of the CDet and CT-INT
  data.}
  \label{fig:sigma_w0_b5_mu068_compare}
\end{figure*}

\subsection{Density of states $A_{loc}(w)$}
We now turn to the density of states (DOS), which provides a momentum-averaged view of the same physics discussed above for $A(\omega,\mathbf{k})$. The DOS is defined as
\begin{equation}
\label{eq:dos_def}
A_{\mathrm{loc}}(\omega) = \frac{1}{N}\sum_{\mathbf{k}} A(\omega,\mathbf{k}),
\end{equation}
and is obtained by analytic continuation of the Matsubara Green's function to the real-frequency axis using the same procedure~\cite{Anacont}. Unlike the momentum-resolved spectra, the DOS emphasizes robust integrated features: the presence (or suppression) of low-energy spectral weight at the Fermi level and the redistribution of spectral weight between low-energy states and the high-energy Hubbard bands.

Figure~\ref{fig:dos_wall} shows $A_{\mathrm{loc}}(\omega)$ obtained within DMFT, fDMFT, and fDF, and compared to the DQMC reference data. The upper panel corresponds to the undoped $\mu = 0.00$ case, while the lower panel shows the doped case $\mu = -0.68$.

In the undoped case (upper panel), paramagnetic DMFT exhibits a pronounced quasiparticle peak at $\omega=0$ for all temperatures shown, indicating a metallic-like solution. In contrast, both fDMFT and fDF display a strong suppression of the low-energy weight and, importantly, no quasiparticle peak at $\omega=0$ at any $\beta$, consistent with the opening of a pseudogap due to strong antiferromagnetic correlations. For intermediate temperatures, $\beta=3$ and $4$, fDMFT and fDF already reproduce the qualitative DQMC trend reasonably well, although the quantitative agreement is not perfect. In this regime, fDF is systematically closer to the DQMC than fDMFT, reflecting the improved treatment of nonlocal correlations. At these temperatures, both fDMFT and fDF develop additional shoulder-like features that are not apparent in the DQMC spectra. Such fine structures should be interpreted with care: the DOS is particularly sensitive to analytic continuation, and residual approximations may generate or enhance weak shoulders. Upon further cooling to $\beta=5$, the agreement becomes significantly better, and the overall DOS profile approaches the DQMC reference much more closely.

The doped case (lower panel) shows a qualitatively different situation. Here the detailed DOS profiles obtained by different approaches are not so closely; nevertheless, the overall temperature trends remain consistent. A crucial limitation is that DQMC in the doped regime suffers from a severe fermionic sign problem, leading to strongly increased statistical noise. This noise is especially detrimental for analytically continued quantities such as $A_{\mathrm{loc}}(\omega)$, where uncertainties on the Matsubara axis can be amplified on the real-frequency axis. Therefore, in the doped case it is difficult to unambiguously assess which fine spectral features are physical, and the comparison should focus primarily on robust qualitative tendencies rather than on small-scale details.

\section{Discussion}
\subsection{Comparison with other methods}

To independently validate our approach, we benchmark it against data from a recent study that performs a cross-comparison between determinant diagrammatic Monte Carlo (CDet) and the interaction-expansion continuous-time quantum Monte Carlo (CT-INT) for the 2D Hubbard model. In direct correspondence with that work, we restrict ourselves to an $8\times 8$ lattice with periodic boundary conditions and compare the momentum-resolved self-energy at the lowest Matsubara frequency, $\Sigma(\mathbf{k}, i\omega_0)$, along the standard high-symmetry path in the Brillouin zone. The model parameters are $U = 5t$,  $\beta=5$ and $\mu = -0.68$ $(\delta n = 0.06)$.  

CT-INT is an established non-perturbative method: it does not require any series resummation (the underlying diagrammatic series converges), but in practice it becomes severely limited by the fermionic sign problem upon doping and as the system size increases (the sign problem vanishes at half-filling). In contrast, CDet evaluates coefficients of the diagrammatic expansion and relies on a resummation procedure, which introduces an additional systematic uncertainty on top of statistical errors, yet enables access to parameter regimes where direct QMC sampling is more challenging.

Overall, we find good agreement between our results and the CDet/CT-INT data on the $8\times 8$ lattice: our curves are consistent with the reported data and accurately reproduce the qualitative momentum dependence of both $\mathrm{Re}\,\Sigma_{\mathbf{k}}(i\omega_0)$ and $\mathrm{Im}\,\Sigma_{\mathbf{k}}(i\omega_0)$. Importantly, the agreement is particularly strong in the vicinity of the Fermi surface, where our method captures not only the correct magnitude but also the characteristic shape and evolution of the self-energy observed in the diagrammatic QMC benchmarks.
\begin{figure}[htb]
  \centering
  \includegraphics[width=\linewidth]{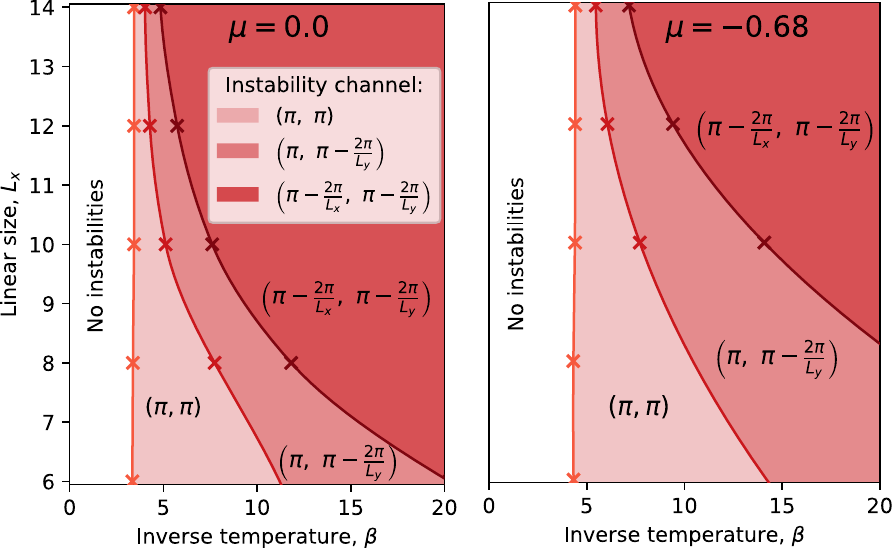}
  \caption{Regions of parameter space of lattice size $L$ and inverse temperature $\beta$ showing the leading instability. Left: undoped $\mu = 0$; right: doped $\mu = -0.68$.}
  \label{fig:leading_channels}
\end{figure}
\subsection{Applicability of the one-channel approach}
We note that the approach described above retains only a single collective fluctuation mode in  the antiferromagnetic channel $s$. In principle, the fDMFT (fDF) approach can be systematically extended to include additional channels and leading fluctuation modes. Here, we delineate the parameter regime where the one-channel approximation is applicable and introduce a practical criterion to assess it in our calculations.

To delineate the regions of parameter space where antiferromagnetic (AF) fluctuations dominate and are only weakly affected by competing channels, we monitor the largest eigenvalue $\Lambda$ of the Bethe-Salpeter kernel $K_{\bf k}$. 
\begin{equation}
K^{\cal{D}}_{\bf k} (\omega_1, \omega_2) = \frac{1}{\beta N} \sum_{\bf q}  \gamma^{\cal{I}}_{\omega_1, \omega_2}  G^{\cal{D}}_{\omega_1, {\bf k}} G^{\cal{D}}_{\omega_2, {\bf k}+{\bf q}},
\label{eq:K_k}
\end{equation}
where $\Gamma^{\cal{I}}_{\omega_1, \omega_2}$ denotes the impurity (irreducible) vertex function. The divergence (instability) criterion is $\Lambda>1$.

As can be seen from Fig.~\ref{fig:leading_channels}, for the cluster sizes relevant for this work ($L=8$-$10$) the leading instability is governed by AF fluctuations in the vicinity of $\mathbf Q=(\pi,\pi)$ and remains well separated from the subleading nearby modes over a broad temperature window. This observation motivates restricting the analysis to moderately small lattices: it maximizes the parameter range where the one-channel (AF) approximation is justified, whereas for larger $L$ and/or lower temperatures the leading mode may shift to the nearest incommensurate momenta, requiring an explicit multi-mode (or multi-channel) treatment.
\subsection{Multi-channel expansion}
Despite the fact that in this work we considered only a fluctuating field $\bf h$ conjugate to the antiferromagnetic spin mode at ${\bf Q}=(\pi,\pi)$, the fDMFT framework is not restricted to a single ordering mode or even to the spin channel. A natural extension of our approach as can be seen in Fig.~\ref{fig:leading_channels} is to include a set of nearby momenta ${\bf q} \approx \left( \pi, \pi \right)$; this should improve the description over a broader range of temperatures and lattice sizes, in particular in regimes where near-commensurate or weakly incommensurate contributions become relevant.

Moreover, in more complex systems the dominant collective modes may arise in other channels as well as the charge (density) modes, where analogous treatments have been explored in the article (\cite{stepanov_FLF2023}). An important practical advantage of our approach is that it can be generalized to such multi-channel with only minor modifications by introducing additional fluctuating fields coupled to the corresponding collective modes. The trade-off is an increased computational cost: each additional mode increases the dimensionality of the auxiliary-field integration, so the numerical effort grows rapidly with the number of included modes.

\section{Conclusion and outlook}
\label{sec:conclusion}
In this work we developed and benchmarked a novel method for describing strongly correlated systems in the intermediate-to-strong coupling regime, focusing on the two-dimensional Hubbard model at $U\approx 5t$ as a prototypical example.

We refer to this approach as fluctuating Dynamical Mean-field Theory (fDMFT). The main idea of fDMFT is to combine the accurate treatment of local correlation physics inherent to conventional DMFT with a consistent description of strong collective fluctuations around the leading nonlocal mode, incorporated through the Fluctuating Local Field (FLF) approach. In this way, the method provides a controlled and physically transparent route to include the contribution of dominant long-range correlations to pure and local physics of DMFT-based schemes.

A key advantage of the proposed approach is its universality: in principle, any DMFT extension could be applied, offering a systematic way to augment it for each desirable system. As a concrete realization of this idea, in the present work, we further enriched fDMFT by incorporating the leading Dual Fermion self-energy diagram (fDF), which accounts for short-range nonlocal correlations beyond the pure local DMFT description and improves the overall agreement with benchmark data.

The resulting scheme enables the computation of key quantities such as the self-energy $\Sigma_{\bf k}(i \omega)$, the spectral function $A_{\bf k}(\omega)$, and the Curie constant $C (\beta)$. Where reference data are available, our results are of comparable accuracy. Even more importantly, the fDMFT approach provides straightforward access to doped regimes, avoiding the difficulties that typically limit QMC simulations away from half-filling.

More generally, an important advantage of the fDMFT framework is its versatility combined with computational simplicity. In practice, it only requires solving the underlying impurity problem for a few dozen values of the fluctuating field $\bf{h}$, which keeps the overall cost moderate and the numerical data essentially noise-free compared to DQMC especially in the doped regime and at low temperatures, where QMC methods are typically hindered by severe statistical noise and/or the sign problem. We also found that adding the leading dual-fermion self-energy contribution (fDF) does not qualitatively alter the physical picture captured at the fDMFT level. This observation suggests that fDMFT alone can provide a reliable and physically plausible description in more complex settings such as multi-orbital models or realistic material Hamiltonians, where diagrammatic corrections rapidly become prohibitively computational expensive. 

At the same time, the present implementation is based on a single leading fluctuation mode, and its applicability is therefore restricted to the region of parameters where this mode is well separated from other nearby collective modes. A natural extension is to generalize the approach by including several competing fluctuation channels, which would substantially broaden the range of accessible regimes. In this direction, a particularly interesting next step is to address superconducting fluctuations in the $t$-$t'$ Hubbard model, relevant to cuprates.

\section*{Acknowledgements}
 S.D.S work was supported by the Foundation for the Advancement of Theoretical Physics and Mathematics ''BASIS'' Grant No. 25-1-5-57-1. A.I.L. supported in part by the ERC Synergy Grant 854843 - FASTCORR.

\bibliography{FLF.bib}
\end{document}